\documentclass[12pt]{article}
\author{A.V.\,Samsonov }
\title{Forward scattering amplitude\\ of the virtual longitudinal
photon in QED}

\begin{document}
\date{}
\maketitle
\it
$$\centerline{\hbox{Institute of Theoretical and Experimental
Physics}}$$
$$\centerline{\hbox{B. Cheremushkinskaya 25, 117259, Moscow, Russia}}$$
  \\ \\

\bf
$$\centerline{\hbox{Abstract}}$$
\rm
\indent Forward scattering amplitude of the virtual longitudinal
photon at zero energy on electron in QED in the limit of small
photon virtualities  $Q^2$ is calculated. The first radiation corrections
are taken into account. Two terms in the expansion over $Q^2$ are obtained.
\\ \\ \\
\large{
\indent B.L.\,Ioffe [1] proved the theorem, determining the forward 
scattering
amplitude of the virtual longitudinal photon at zero energy on any target
in the limit of small photon virtualities $Q^2$. The only assumption about
interaction was the presence of a gap in the mass spectrum. It was shown     
that two terms in the expansion of the amplitude over $Q^2$ are universal
and expressed in terms of static target properties. It is interesting to
study, if such theorem is valid in the case where is no gap in the mass
spectrum. 
An example of the theory with no gap in the mass spectrum is QED.\\
\indent In this paper the forward scattering amplitude of the virtual 
longitudinal 
photon on electron at zero energy is calculated in QED at order $\alpha^2$
in the limit of small $Q^2$ and two terms of the expansion over $Q^2$ are
retained. It was found that in the leading $Q^2$ term the universality
indeed takes place and the theorem [1] is valid, but in the next 
$Q^2$ term it is violated; i.e. the results in presence and 
absence of the mass gap are different. \\
\indent For real (transverse) photon the low energy theorem for Compton
scattering was proved by Thirring [2] and Kroll and Ruderman [3];
it gives Thomson formula for the amplitude: $f=-\alpha /m\;$.\\
\\ \indent Now we consider the forward scattering amplitude of the virtual
longitudinal photon at zero energy on electron in QED. We assume that
$Q^2$ tends to zero and calculate all non-vanishing terms at order $e^4$.
This amplitude is the function of two
invariants $\nu=pq$ and $q^2$, where $p$ and $q$ are the electron and
photon 4-momenta. Our conditions are:$$ \nu=pq=0\;,\;\;\;\;\;\;
q^2<0\;,\;\;\;\;\;\;q^2 \to 0\;.\\$$
We use metric $(1,-1,-1,-1)$.
We can choose coordinate system, where $z$ axis is the collision one and
$q_0=0$. Then $ q^2=-{q_z}^2$,$\; q_\mu=(0,0,0,q_z)\;$ and
 $p_\mu=(m,0,0,0)$. If $e_\mu$ is the photon polarization vector, then
 $e_\mu q_\mu=0\;$ and $e_\mu=(1,0,0,0)$, because we consider longitudinal
 photon.\\
\indent At first we find our amplitude in the tree approximation $M_0$ 
(Fig 1).\\
$$M_0={1\over{2im}} \overline{u}(p) \gamma_0 e_0 (-ie)
({i\over{\hat p+\hat q-m}}+{i\over{\hat p-\hat q-m}})
 (-ie) e_0 \gamma_0 u(p)=-{e^2\over{mq^2}}\ 4m^2\;,$$
or, introducing $Q^2=-q^2$,
$$M_0=4{me^2\over{Q^2}}\;.$$
It is easy to see singular behaviour of the amplitude when $Q^2$ becomes
zero. The sign of the amplitude is opposite to one in the Thomson formula.\\
\indent Now we consider changes from  diagrams with radiation corrections
of the first order. Let us first in all expressions retain $1/{q^2}$ terms
only. \\
\indent 1. Diagrams with mass operator $M_1$ (Fig 2).\\
Exact electron propogator $ {\hat G}_r(p^\prime) $ has the form:
$${\hat G}_r(p^\prime)={1\over{{\hat p}^\prime-m-{\hat \Sigma}_r
(p^\prime)}}\;\;\;, \eqno(1) $$
where $ p^\prime=p+q $, and
$ {\hat \Sigma}_r(p^\prime) $ -exact mass operator
$$ {\hat \Sigma}_r(p^\prime)={{1\over{1-{{\hat \Sigma}^\prime}(m)}}
[\;{\hat \Sigma}(p^\prime)-{\hat \Sigma}(m)
-({\hat p}^\prime-m){{\hat \Sigma}^\prime}(m)\;]}\;, $$
where
$$ {{\hat \Sigma}^\prime}(m)=({{\partial{{\hat \Sigma}(p^\prime)}
/ \partial{p^\prime}})\Big{\vert}_{p^\prime=m}}\;\;\;. $$
Therefore:
$$ {\hat G}_r(p^\prime)={{1-{{\hat \Sigma}^\prime}(m)}\over{
{\hat p}^\prime-m-[\;{\hat \Sigma}(p^\prime)-{\hat \Sigma}(m)\;]}\;\;}\;. $$
Using expansion
$$ {\hat \Sigma}(p^\prime)-{\hat \Sigma}(m)=
({\hat p}^\prime-m)\>{{\hat \Sigma}^\prime}(m)+
({\hat p}^\prime-m)^2\>{{\hat \Sigma}^{\prime\prime}}(m)+...\;\;\;,$$
we obtain for $ {\hat G}_r(p^\prime)\;\; $:
$$ {\hat G}_r(p^\prime)={1\over{\hat p+\hat q-m+b(\hat p+\hat q-m)^2}}
\;\;\;, $$
where b - constant, which comes out of the contributions from
radiation corrections when $q_z \to 0$.\\
But
 $$ (\hat p+\hat q-m)^2=2m^2+q^2-2m\>(\hat p+\hat q)\;,$$
because $\nu=0$ , and therefore (1) has the following form:
$$ {\hat G}_r(p^\prime)={1\over{(\hat p+\hat q)(1-2mb)-m(1-2mb)+bq^2}}
\;\;.$$
Substituting this expression (and the expression for another diagram,
for which we have to replace $q_\mu$ by $-q_\mu$ ) and retaining $1/q^2$
terms only, we obtain ($ \overline{u} \hat q u=0 $):
$$M_1=-{e^2\over{m}}\overline{u}(p)\>{{(\hat p+\hat q+m)(1-2mb)}\over
{q^2\,(1-2mb)^2+2mbq^2\,(1-2mb)}}\>u(p)=M_0\;.$$
We see that radiation corrections to propogator (i.e. constant b)
don't give contribution (proportional to $1/q^2$) to the amplitude $M_0$.\\
\indent 2. Vertex diagrams  $M_2$ (Fig 3).\\
Vertex has the form:
$${\Gamma_r}_\mu=\gamma_\mu+b(\gamma_\mu \hat q-\hat q\gamma_\mu)\;.$$
Using that $e_\mu={1\over{m}}p_\mu\;$, we obtain:
$$\displaylines{ M_2=-{e^2\over{m}}{\overline{u}(p)\> {\hat p \over{m}}
{1\over{(\hat p+\hat q-m)}}\>{1\over{m}}(\hat p+2b\hat p\hat q)\>u(p)}-
\hfill \cr} $$
$$\displaylines{\hfill -{e^2\over{m}}{\overline{u}(p)\> {\hat p \over{m}}
{1\over{m}}(\hat p+2b\hat p\hat q)\>{1\over{(\hat p+\hat q-m)}}\>u(p)}=
M_0\;. \cr }$$
Just as in the first case, constant b vanishes from the answer. So vertex
corrections don't give additional contribution to $M_0$.\\
\indent 3. Box diagrams $M_3$ (Fig 4).\\
If $\hat G(p)\;$- electron propogator: $\hat G(p)=1/(\hat p-m)\;,$
then ${\partial{\hat G(p)}/{\partial{p_\mu}}}$
is the vertex with zero photon momentum. Therefore we obtain two $M_3$
diagrams (in the limit $q^2 \to 0 $) after double differentiation of 
mass operator with respect to incoming momentum. It is easy to see
that this expression has no singularity $1/q^2$.\\
\indent 4. Diagrams with vacuum polarization $M_4$ (Fig 5).\\
Polarization operator has the following form ( $- q^2<<m^2$):
$$\Pi_{\mu\nu}=\Pi(q^2)\,(g_{\mu\nu}-{{q_\mu q_\nu}\over{q^2}})\;,
\;\;\;\; \Pi(q^2)=-{e^2\over{15\pi}}\,{q^4\over{m^2}}\;\;.$$
That's why $M_4$ diagrams don't contain $1/q^2$ terms.\\
\indent As a result we can conclude that  the forward scattering amplitude
of the photon with $q_\mu=(0,0,0,q_z)$ and $e_\mu=(1,0,0,0)$ on 
electron with $p_\mu=(m,0,0,0)$, being calculated with the first radiation
corrections, has the form:
$$M=4{me^2\over Q^2}+R\;,\eqno(2) $$
where $Q^2=-q^2$, and $R\;$ contains non-negative degrees of $Q^2.$\\ 
\indent In paper [1] for such amplitude was obtained the following expression
(assuming the presence of the gap):
$$f_L(0,Q^2)={4e^2m\over{Q^2}}\,[1-{Q^2\over{2m^2}}\mu_a-{1\over 3}\,
\langle r^2_E \rangle \, Q^2]\;,\eqno(3) $$
where $\mu_a$ and $\langle r^2_E \rangle \,$ are anomalous magnetic
moment and target mean square radius. 
We see that the first terms of expansion over $Q^2$ coincide in equations
(2) and (3).\\
\indent In order to obtain the second term of the expansion of the amplitude
over $Q^2$ (i.e. constant R in (2)), we have to calculate directly all
above-mentioned diagrams. \\
Let us introduce two new parameters:
$$\rho=-q^2/m^2\;,\;s=\lambda^2/m^2\;,$$
where $\lambda$ is the photon mass. \\
We assume that
$$ 0<s<<\rho<<1\;,\;\;\;\;\;\;\;\;\;\rho \to 0\;.\eqno(4) $$
Under such conditions we obtain:

$$\displaylines{M_1={e^4\over{\pi m}}[\int_0^1{dx\,{4\over \rho^2}\,(1+x)\,
\ln{{x^2+\rho x(1-x)+s(1-x)}\over{x^2+s(1-x)}}}\>+\hfill\cr}$$
$$\displaylines{\hfill+\int_0^1{dx\,{1\over \rho}\,((1-3x)\,
\ln{{x^2+\rho x(1-x)+s(1-x)}\over{x^2+s(1-x)}}\>-
{{4x(1-x^2)}\over{x^2+s(1-x)}})}]=\cr}$$
$$\displaylines{ ={e^4\over {\pi m}}({\,4 \over \rho}
-{4 \over \rho}\, ln \rho +{2 \over \rho}\, ln s -3\, ln \rho
+{1 \over 2}\,)\;.\hfill\cr} $$

$$\displaylines{M_2=2{e^4\over{\pi m\rho}}\int_0^1{dx\int_x^1
{dy\,[{{2(y^2+2y-2)+
\rho\,(2x^2-1+y-2xy)}\over{y^2+\rho x(1-x)+s\,(1-y)}}}}\;+\hfill\cr}$$
$$\displaylines{\hfill +2\ln{{y^2+s\,(1-y)}\over
{y^2+\rho x(1-x)+s\,(1-y)}}-{{2(y^2+2y-2)}\over{y^2+s\,(1-y)}}]=\cr}$$
$$\displaylines{ ={e^4\over {\pi m}}(\,-{ 8 \over \rho}
+{8 \over \rho}\, ln \rho -{4 \over \rho}\, ln s +6\, ln \rho
-{277 \over 15}\,)\;.\hfill\cr} $$

$$\displaylines{M_3=-{e^4\over{\pi m}}\int_0^1{dx\int_x^1{dy\,(y-x)
({{y^3+3y^2-6y+4+
\rho\,(3yx^2-x^2-2x-2xy)}\over{(y^2+\rho x(1-x)+s\,(1-y))^2}}}}-\hfill\cr}$$
$$\displaylines{\hfill -{{3-3y}\over{y^2+\rho x(1-x)+s\,(1-y)}})= \cr}$$
$$\displaylines{ ={e^4\over {\pi m}}({\,4 \over \rho}
-{4 \over \rho}\, ln \rho +{2 \over \rho}\, ln s -3\, ln \rho
+{11 \over 6}\,)\;.\hfill\cr} $$

$$\displaylines{ M_4={8 \over 15}\,{ e^4\over {\pi m}}\;.\hfill\cr} $$

In this expressions we retain only singular terms and constants. 
$$R=M_1+M_2+M_3+M_4=-{78\over{5}}{e^4\over{\pi m}}\;.$$
\indent The final result for the forward scattering amplitude $M$ has the
following form:
$$M=4{me^2\over Q^2}-{78\over{5}}{e^4\over{\pi m}}\;.\eqno(5) $$
According to the first part of this article, radiation
corrections diagrams  don't give any contribution to the singular term.\\
\indent The next to leading term in the expansion over $Q^2$ of the 
amplitude (3) in QED is equal to $\;{4/3}\,({e^4/{\pi m}})\,ln s.\;$ 
(The expressions for $\mu_a$ and
$\langle r^2_E \rangle \,$ at order $e^4$ are substituted from [4]\,).
That's why (3) at order $e^4$ has the following form:
$$f_L(0,Q^2)=4{me^2\over{Q^2}}+
{4\over 3}\,{e^4\over{\pi m}}\,ln s \;. \eqno(6) $$
The second terms in (5) and (6) don't coincide. Expression (5)
contains neither photon mass explicitly (it has been obvious from the
beginning) nor experimental resolution $\Delta \epsilon$.\\
\indent It can be shown that formula (6) is reproduced in
the case of the finite mass gap, which in our notation corresponds to
condition
$$ 0<\rho<<s<<1\;,\;\;\;\;\;\;\;\;\;s \to 0\;.$$

\indent In all calculations $\nu$ equals zero, and $\lambda$ equals 
 very small but nonzero quantity. But if we
 put nonzero $\nu$ and change condition (4) to the following one:
$$ 0<s<<{\nu\over m^2}<<\rho<<1\;,\;\;\;\;\;\;\;\rho \to 0\;,
\;\;\;\;\nu \to 0\;,$$
we obtain the same answer (5) for the amplitude $M$.\\ \\
\indent I am grateful to prof. B.L.\,Ioffe for formulation of the problem 
and helpful discussions.
\\ \\ \\ \\
\centerline{\Large{References.}}\\ \\
$[1]$ B.L.\,Ioffe Phys.Rev.D 55 R1130 (1997).\\
$[2]$ W.Tirring, Phil.Mag. 41, 1193 (1950).\\
$[3]$ N.Kroll, M.Ruderman, Phys.Rev. 93, 233 (1954).\\
$[4]$ V.Berestetskii, E.Lifshitz, L.Pitaevskii 'Quantum electrodynamics',
 \\ 'Nauka', 1989.
\newpage
\centerline{\Large{Figures.}} $\;$ \\  $\;$
\unitlength=1.00mm
\special{em:linewidth 0.4pt}
\linethickness{0.4pt}
\begin{picture}(66.00,50.00)
\bezier{48}(30.00,35.00)(24.00,33.00)(27.00,38.00)
\bezier{48}(50.00,35.00)(56.00,33.00)(53.00,38.00)
\bezier{40}(53.00,38.00)(51.00,42.00)(56.00,41.00)
\bezier{40}(27.00,38.00)(29.00,42.00)(24.00,41.00)
\bezier{40}(24.00,41.00)(20.00,39.00)(21.00,44.00)
\bezier{40}(56.00,41.00)(60.00,39.00)(59.00,44.00)
\put(15.00,15.00){\vector(3,4){15.00}}
\put(50.00,35.00){\vector(3,-4){15.00}}
\put(55.00,20.00){\makebox(0,0)[cc]{p}}
\put(25.00,20.00){\makebox(0,0)[cc]{p}}
\bezier{48}(59.00,44.00)(56.00,49.00)(62.00,47.00)
\bezier{48}(21.00,44.00)(24.00,49.00)(18.00,47.00)
\bezier{40}(18.00,47.00)(14.00,45.00)(15.00,50.00)
\bezier{40}(62.00,47.00)(66.00,45.00)(65.00,50.00)
\put(50.00,45.00){\makebox(0,0)[cc]{q}}
\put(30.00,45.00){\makebox(0,0)[cc]{q}}
\put(30.00,35.00){\vector(1,0){20.00}}
\put(40.00,10.00){\makebox(0,0)[cc]{Figure 1}}
\end{picture}
\unitlength=1.00mm
\special{em:linewidth 0.4pt}
\linethickness{0.4pt}
\begin{picture}(76.00,50.00)
\put(20.00,35.00){\vector(1,0){10.00}}
\put(50.00,35.00){\vector(1,0){10.00}}
\bezier{44}(30.00,35.00)(32.00,30.00)(34.00,35.00)
\bezier{44}(34.00,35.00)(36.00,40.00)(38.00,35.00)
\bezier{44}(38.00,35.00)(40.00,30.00)(42.00,35.00)
\bezier{44}(42.00,35.00)(44.00,40.00)(46.00,35.00)
\bezier{44}(46.00,35.00)(48.00,30.00)(50.00,35.00)
\bezier{48}(20.00,35.00)(14.00,33.00)(17.00,38.00)
\bezier{48}(60.00,35.00)(66.00,33.00)(63.00,38.00)
\bezier{40}(63.00,38.00)(61.00,42.00)(66.00,41.00)
\bezier{40}(17.00,38.00)(19.00,42.00)(14.00,41.00)
\bezier{40}(14.00,41.00)(10.00,39.00)(11.00,44.00)
\bezier{40}(66.00,41.00)(70.00,39.00)(69.00,44.00)
\put(5.00,15.00){\vector(3,4){15.00}}
\put(60.00,35.00){\vector(3,-4){15.00}}
\put(40.00,34.50){\oval(20.00,19.00)[t]}
\put(65.00,20.00){\makebox(0,0)[cc]{p}}
\put(15.00,20.00){\makebox(0,0)[cc]{p}}
\bezier{48}(69.00,44.00)(66.00,49.00)(72.00,47.00)
\bezier{48}(11.00,44.00)(14.00,49.00)(8.00,47.00)
\bezier{40}(8.00,47.00)(4.00,45.00)(5.00,50.00)
\bezier{40}(72.00,47.00)(76.00,45.00)(75.00,50.00)
\put(60.00,45.00){\makebox(0,0)[cc]{q}}
\put(20.00,45.00){\makebox(0,0)[cc]{q}}
\put(40.00,10.00){\makebox(0,0)[cc]{Figure 2}}
\end{picture}
 \\
\unitlength=1.00mm
\special{em:linewidth 0.4pt}
\linethickness{0.4pt}
\begin{picture}(142.00,55.00)
\bezier{48}(36.00,40.00)(30.00,38.00)(33.00,43.00)
\bezier{48}(34.00,23.00)(40.00,21.00)(37.00,26.00)
\bezier{40}(37.00,26.00)(35.00,30.00)(40.00,29.00)
\bezier{40}(33.00,43.00)(35.00,47.00)(30.00,46.00)
\bezier{40}(30.00,46.00)(26.00,44.00)(27.00,49.00)
\bezier{40}(40.00,29.00)(44.00,27.00)(43.00,32.00)
\bezier{48}(43.00,32.00)(40.00,37.00)(46.00,35.00)
\bezier{48}(27.00,49.00)(30.00,54.00)(24.00,52.00)
\bezier{40}(24.00,52.00)(20.00,50.00)(21.00,55.00)
\bezier{40}(46.00,35.00)(50.00,33.00)(49.00,38.00)
\put(56.00,50.00){\makebox(0,0)[cc]{q}}
\put(36.00,50.00){\makebox(0,0)[cc]{q}}
\put(36.00,40.00){\vector(1,0){20.00}}
\put(30.00,25.00){\vector(1,3){5.00}}
\bezier{48}(56.00,40.00)(62.00,38.00)(59.00,43.00)
\bezier{40}(59.00,43.00)(57.00,47.00)(62.00,46.00)
\bezier{40}(62.00,46.00)(66.00,44.00)(65.00,49.00)
\bezier{48}(65.00,49.00)(62.00,54.00)(68.00,52.00)
\bezier{40}(68.00,52.00)(72.00,50.00)(71.00,55.00)
\bezier{16}(49.00,38.00)(48.00,39.00)(50.00,40.00)
\bezier{20}(34.00,23.00)(32.00,23.00)(30.00,25.00)
\put(20.00,15.00){\vector(1,1){10.00}}
\put(56.00,40.00){\vector(2,-3){16.67}}
\put(74.00,20.00){\makebox(0,0)[cc]{p}}
\put(20.00,20.00){\makebox(0,0)[cc]{p}}
\bezier{48}(106.00,40.00)(100.00,38.00)(103.00,43.00)
\bezier{40}(103.00,43.00)(105.00,47.00)(100.00,46.00)
\bezier{40}(100.00,46.00)(96.00,44.00)(97.00,49.00)
\bezier{48}(97.00,49.00)(100.00,54.00)(94.00,52.00)
\bezier{40}(94.00,52.00)(90.00,50.00)(91.00,55.00)
\put(126.00,50.00){\makebox(0,0)[cc]{q}}
\put(106.00,50.00){\makebox(0,0)[cc]{q}}
\put(106.00,40.00){\vector(1,0){20.00}}
\bezier{48}(126.00,40.00)(132.00,38.00)(129.00,43.00)
\bezier{40}(129.00,43.00)(127.00,47.00)(132.00,46.00)
\bezier{40}(132.00,46.00)(136.00,44.00)(135.00,49.00)
\bezier{48}(135.00,49.00)(132.00,54.00)(138.00,52.00)
\bezier{40}(138.00,52.00)(142.00,50.00)(141.00,55.00)
\put(126.00,40.00){\vector(1,-3){5.00}}
\put(131.00,25.00){\vector(1,-1){10.00}}
\put(88.00,15.00){\vector(2,3){16.67}}
\bezier{48}(128.00,23.00)(122.00,21.00)(125.00,26.00)
\bezier{40}(125.00,26.00)(127.00,30.00)(122.00,29.00)
\bezier{40}(122.00,29.00)(118.00,27.00)(119.00,32.00)
\bezier{48}(119.00,32.00)(122.00,37.00)(116.00,35.00)
\bezier{40}(116.00,35.00)(112.00,33.00)(113.00,38.00)
\bezier{16}(128.00,23.00)(129.00,23.00)(131.00,25.00)
\bezier{12}(113.00,38.00)(112.00,40.00)(113.00,40.00)
\put(86.00,20.00){\makebox(0,0)[cc]{p}}
\put(141.00,20.00){\makebox(0,0)[cc]{p}}
\put(80.00,10.00){\makebox(0,0)[cc]{Figure 3}}
\end{picture}
\\ 
\unitlength=1.00mm
\special{em:linewidth 0.4pt}
\linethickness{0.4pt}
\begin{picture}(105.00,50.00)
\bezier{44}(66.00,20.00)(68.00,15.00)(70.00,20.00)
\bezier{44}(70.00,20.00)(72.00,25.00)(74.00,20.00)
\bezier{44}(74.00,20.00)(76.00,15.00)(78.00,20.00)
\bezier{44}(78.00,20.00)(80.00,25.00)(82.00,20.00)
\bezier{44}(82.00,20.00)(84.00,15.00)(86.00,20.00)
\bezier{48}(71.00,35.00)(65.00,33.00)(68.00,38.00)
\bezier{48}(89.00,35.00)(95.00,33.00)(92.00,38.00)
\bezier{40}(92.00,38.00)(90.00,42.00)(95.00,41.00)
\bezier{40}(68.00,38.00)(70.00,42.00)(65.00,41.00)
\bezier{40}(65.00,41.00)(61.00,39.00)(62.00,44.00)
\bezier{40}(95.00,41.00)(99.00,39.00)(98.00,44.00)
\bezier{48}(98.00,44.00)(95.00,49.00)(101.00,47.00)
\bezier{48}(62.00,44.00)(65.00,49.00)(59.00,47.00)
\bezier{40}(59.00,47.00)(55.00,45.00)(56.00,50.00)
\bezier{40}(101.00,47.00)(105.00,45.00)(104.00,50.00)
\put(89.00,45.00){\makebox(0,0)[cc]{q}}
\put(71.00,45.00){\makebox(0,0)[cc]{q}}
\bezier{44}(86.00,20.00)(88.00,25.00)(90.00,20.00)
\bezier{44}(90.00,20.00)(92.00,15.00)(94.00,20.00)
\put(66.00,20.00){\vector(1,3){5.00}}
\put(89.00,35.00){\vector(1,-3){5.00}}
\put(71.00,35.00){\vector(1,0){18.00}}
\put(56.00,10.00){\vector(1,1){10.00}}
\put(94.00,20.00){\vector(1,-1){10.00}}
\put(56.00,15.00){\makebox(0,0)[cc]{p}}
\put(104.00,15.00){\makebox(0,0)[cc]{p}}
\put(80.00,5.00){\makebox(0,0)[cc]{Figure 4}}
\end{picture}
\\ 
\unitlength=1.00mm
\special{em:linewidth 0.4pt}
\linethickness{0.4pt}
\begin{picture}(146.00,50.00)
\bezier{48}(40.00,25.00)(34.00,23.00)(37.00,28.00)
\bezier{48}(60.00,25.00)(66.00,23.00)(63.00,28.00)
\bezier{40}(63.00,28.00)(61.00,32.00)(66.00,31.00)
\bezier{40}(37.00,28.00)(39.00,32.00)(34.00,31.00)
\bezier{40}(34.00,31.00)(30.00,29.00)(31.00,34.00)
\bezier{40}(66.00,31.00)(70.00,29.00)(69.00,34.00)
\put(65.00,10.00){\makebox(0,0)[cc]{p}}
\put(35.00,10.00){\makebox(0,0)[cc]{p}}
\bezier{48}(69.00,34.00)(66.00,39.00)(72.00,37.00)
\bezier{48}(22.00,44.00)(25.00,49.00)(19.00,47.00)
\bezier{40}(19.00,47.00)(15.00,45.00)(16.00,50.00)
\bezier{40}(72.00,37.00)(76.00,35.00)(75.00,40.00)
\put(60.00,35.00){\makebox(0,0)[cc]{q}}
\put(40.00,25.00){\vector(1,0){20.00}}
\bezier{48}(101.00,25.00)(95.00,23.00)(98.00,28.00)
\bezier{48}(121.00,25.00)(127.00,23.00)(124.00,28.00)
\bezier{40}(124.00,28.00)(122.00,32.00)(127.00,31.00)
\bezier{40}(98.00,28.00)(100.00,32.00)(95.00,31.00)
\bezier{40}(95.00,31.00)(91.00,29.00)(92.00,34.00)
\bezier{40}(127.00,31.00)(131.00,29.00)(130.00,34.00)
\put(126.00,10.00){\makebox(0,0)[cc]{p}}
\put(96.00,10.00){\makebox(0,0)[cc]{p}}
\bezier{48}(139.00,43.00)(136.00,48.00)(142.00,46.00)
\bezier{48}(92.00,34.00)(95.00,39.00)(89.00,37.00)
\bezier{40}(89.00,37.00)(85.00,35.00)(86.00,40.00)
\bezier{40}(142.00,46.00)(146.00,44.00)(145.00,49.00)
\put(101.00,35.00){\makebox(0,0)[cc]{q}}
\put(101.00,25.00){\vector(1,0){20.00}}
\put(28.00,38.00){\circle{10.20}}
\bezier{40}(22.00,44.00)(19.00,40.00)(24.00,41.00)
\put(133.00,38.00){\circle{10.00}}
\bezier{24}(139.00,43.00)(140.00,40.00)(137.00,41.00)
\put(27.00,48.00){\makebox(0,0)[cc]{q}}
\put(134.00,48.00){\makebox(0,0)[cc]{q}}
\put(80.00,-1.00){\makebox(0,0)[cc]{Figure 5}}
\put(25.00,10.00){\vector(1,1){15.00}}
\put(60.00,25.00){\vector(1,-1){15.00}}
\put(86.00,10.00){\vector(1,1){15.00}}
\put(121.00,25.00){\vector(1,-1){15.00}}
\end{picture}
\newpage
\centerline{\Large{Figure captions.}}$\;$   $\;$\\
Fig. 1 -tree diagrams. \\
Fig. 2 -diagrams with mass operator. \\
Fig. 3 -vertex diagrams. \\
Fig. 4 -box diagrams. \\
Fig. 5 -diagrams with polarization operator.
\end{document}